\documentclass[aps,prl,twocolumn,showpacs,superscriptaddress,floatfix,nofootinbib]{revtex4}

\usepackage{graphicx}
\usepackage{amsmath}
\usepackage{epsfig}
\usepackage{helvet}
\usepackage{amssymb}

\newcommand{\be}{\begin{equation}}
\newcommand{\ee}{\end{equation}}
\newcommand{\bea}{\begin{eqnarray}}
\newcommand{\eea}{\end{eqnarray}}

\newcommand{\tr}{\mbox{tr}}

\newcommand{\ket}[1]{\mbox{$| #1 \rangle$}}

\def\tr{ \mbox{tr}}

\begin{document}

\title{
Entanglement renormalization in fermionic systems}
\author{G. Evenbly}
\author{G. Vidal}
\affiliation{School of Physical Sciences, the University of
Queensland, QLD 4072, Australia} 
\date{\today}

\begin{abstract}
We demonstrate, in the context of quadratic fermion lattice models in one and two spatial dimensions, the potential of entanglement renormalization (ER) to define a proper real-space renormalization group transformation. Our results show, for the first time, the validity of the multi-scale entanglement renormalization ansatz (MERA) to describe ground states in two dimensions, even at a quantum critical point. They also unveil a connection between the performance of ER and the logarithmic violations of the boundary law for entanglement in systems with a one-dimensional Fermi surface. ER is recast in the language of creation/annihilation operators and correlation matrices.
\end{abstract}


\maketitle

The \emph{renormalization group} (RG), concerned with the change of physics with the observation scale, is among the main ideas underlying the theoretical structure of statistical mechanics and quantum field theory, and of central importance in the modern formulation of critical phenomena and phase transitions \cite{fisher}. Its influence extends well beyond the conceptual domain: RG transformations are also the basis of numerical approaches to the study of strongly correlated many-body systems.

In a lattice model, a real-space RG transformation produces a coarse-grained system by first joining the lattice sites into blocks and then replacing each block with an effective site \cite{kadanov}. Two very natural requirements for a such RG transformation are: ($i$) it should preserve the long-distance physics of the system; ($ii$) when this physics is invariant under changes of scale, the system should be a fixed point of the RG transformation. 

For the important case of a quantum system at zero temperature, the first requirement is fulfilled if, as determined by White in his density matrix renormalization group (DMRG) \cite{dmrg}, the vector space of the effective site retains the local support of the ground state. \emph{Entanglement renormalization} (ER) \cite{ER} has recently been proposed in order to simulatenously meet the second requirement. By using \emph{disentanglers}, ER aims to produce a coarse-grained lattice locally identical to the original one, in the sense that their sites have the same vector space dimension. When this is accomplished, the original system and its coarse-grained version can be meaninglfully compared, e.g. through their Hamiltonians or ground state properties, leading to a proper real-space RG flow. 

Promisingly, ER has been successfully demonstrated for the 1D quantum Ising model with transverse magnetic field, where it has been shown that, indeed, at the quantum critical point the system is invariant under the resulting RG transformation \cite{ER}. However, plenty of work is still required to characterize the main features and range of applicability of this new approach and, in particular, it remains to be seen whether ER also works in the computationally more challenging context of 2D lattice systems, where DMRG can no longer analyse large systems.

In this paper we explore the performance of ER in systems of spinless fermion both on 1D and 2D lattices, as specified by the quadratic Hamiltonian
\begin{equation}
 \hat{H} = \sum_{r,s=1}^N {\frac{1}{2}\left[ {\hat a_r^\dag  \hat a_s  + \gamma \left( {\hat a_r^\dag  \hat a_s^\dag   + \hat a_s \hat a_r } \right)} \right]}  - \lambda\sum_{r=1}^N { \hat a_r^\dag  \hat a_r },\label{eq:Ham} 
\end{equation}
where $\lambda$ and $\gamma$ are the chemical and pairing potentials and the first sum involves only nearest neighbors. In spite of its simplicity, Hamiltonian $\hat{H}$ contains a rich phase diagram as a function of $\lambda$ and $\gamma$, including insulating, conducting and superconducting phases \cite{Haas}. Importantly, the corresponding ground states span all known forms of entropy scaling \cite{Haas, Scholl}. 
In addition, $\hat{H}$ can be diagonalized through linear (Fourier and Bogoliubov) transformations of the fermion operators $\hat a$ and $\hat a^\dag$ while, by Wick's theorem, all properties of its \emph{gaussian} ground state $\ket{\Psi_{\mbox{\tiny GS}}}$ can be extracted from the two-point correlators $\left\langle {\hat a_r^\dag  \hat a_s} \right\rangle$ and $\left\langle {\hat a_r  \hat a_s} \right\rangle$. Then, provided that our RG transformation also maps fermion modes linearly, the entire analysis can be conducted in the space of two-point correlators and quadratic Hamiltonians of $N$ fermionic modes, as represented by $N\times N$ matrices. Hence quadratic fermionic models such as (\ref{eq:Ham}) offer an appealing testing ground for ER, one where computational costs have been greatly simplified (e.g. $\hat{H}$ can be diagonalized exactly with just $O(N^3)$ operations) while keeping a rich variety of non-trivial ground state structures. 

We start by rephrasing, in the language of correlation matrices, the process of coarse-graining a $D$-dimensional (hypercubic) lattice. We assume that the system is in the ground state $\ket{\Psi_{\mbox{\tiny GS}}}$ of $\hat{H}$, which we compute using standard analytic techniques (see e.g. \cite{Haas}). 
It is convenient to redraw the hypercubic lattice so that each site contains $P\equiv p^D$ fermion modes for some integer $p$. Then a hypercube of $2^D$ sites defines a \emph{block} that contains $P2^{D}$ modes. The goal of the RG transformation is to replace this block with just one effective site made of $P'$ modes, with $P'<P2^{D}$. We would like to have $P'=P$, so that the sites of the coarse-grained and original lattices are identical and we can compare the corresponding Hamiltonians or ground-state reduced density matrices. However, in the coarse-graining step only modes of the block that are disentangled from the rest of the system can be removed (see appendix A). As a result, $P'$ often must be larger than $P$.  

For the sake of simplicity, we continue the analysis for the case of a 1D lattice (see appendix B for construction deatils of MERA for the 2D lattice). Let us temporarily replace the $N$ spinless fermion operators $\hat a$ in Eq. (\ref{eq:Ham}) with $2N$ (self-adjoint) Majorana fermion operators $\check c$,   
\begin{equation}
\check c_{2r - 1} \equiv \hat a_r + \hat a_r^{\dagger},~~~~~~~~~\check c_{2r} \equiv \frac{\hat a_r - \hat a_r^{\dagger}}{i}. \label{eq:Majorana}
\end{equation}
The ground state $\ket{\Psi_{\mbox{\tiny GS}}}$ is then completely specified by
\begin{equation}
\left\langle {\check c_r \check c_s } \right\rangle  = \delta _{rs}  + i \Gamma_{rs}, \label{eq:Gamma} 
\end{equation}
where $\Gamma$, henceforth referred to as the \emph{correlation matrix}, is real and antisymmetric. Similarly, the reduced density matrix $\rho_{\mbox{\tiny GS}}$ for a block made of 2 sites, that is with $L=2P$ spinless modes (equivalently, $2L$ Majorana modes) is described by a $2L\times 2L$ submatrix $\Gamma_{L}$ of $\Gamma$. This matrix is brought into (block) diagonal form by a special orthogonal transformation $V$,
\begin{equation}
 V\Gamma _L V^\dag = \bigoplus \limits_{r = 1}^L \left[ {\begin{array}{cc}
   0 & {v_r }  \\
   { - v_r } & 0  \\
\end{array}} \right], ~~~~  V \in SO(2L),\label{eq:GammaL} 
\end{equation}
where $0 \le v_r \le 1$ are the eigenvalues of $\Gamma_L$, each one associated with a pair of Majorana fermions. These pairs recombine into $L$ spinless fermions in a product state \cite{logVidal} 
\begin{equation}
	\rho_{\mbox{\tiny GS}} = \bigotimes_{r=1}^L\varrho_{r} = \bigotimes_{r=1}^L \left( {\begin{array}{cc}
   \frac{1+v_r}{2} & 0  \\
   0 & \frac{1-v_r}{2}  \\
\end{array}} \right), \label{eq:rho}
\end{equation}
where $\varrho_r$, the state of a spinless fermion mode, is \emph{mixed} if $v_r<1$ and \emph{pure} if $v_r=1$. Notice that since the ground state $\ket{\Psi_{\mbox{\tiny GS}}}$ is a pure state, a mode in a mixed state must be \emph{entangled} with modes outside the block, whereas a mode in a pure state is \emph{unentangled} from the rest of the system. We build an effective site by removing from the block, or \emph{projecting out} from $\Gamma_L$, all the modes that are unentangled (pure), and just keeping those $P'$ modes that are entangled (mixed). In this way, the coarse-grained lattice retains the ground state properties, see appendix.

\begin{figure}
\begin{center}
\includegraphics[width=8cm]{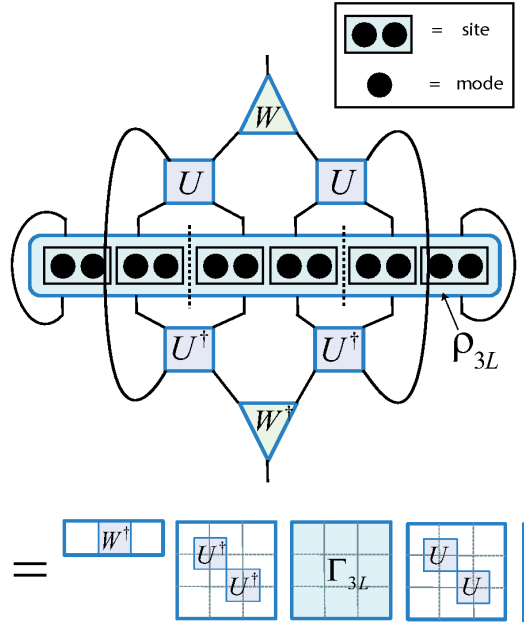}
\caption{\emph{Top:} A block of two sites (four modes) is coarse-grained into an effective site by first applying disentanglers $U$ across the boundary of the block and then using isometry $W$ to project out two modes. \emph{Bottom:} Same RG transformation written in the language of correlation matrices, Eq. (\ref{eq:truncated}).} \label{fig:disentanglers}
\end{center}
\end{figure}

The key idea of ER, see Fig. (\ref{fig:disentanglers}), is to use disentangling unitary transformations, or \emph{disentanglers}, to diminish $P'$ by increasing the number of modes in the block that are unentangled from the rest of the system. A disentangler is implemented through a special orthogonal matrix $U\in SO(2L)$ that acts on two neigboring sites across the boundary of the block,  wheareas the coarse-graining is implemented by an isometry $W = RY_{P'}$ that selects the $P'$ spinless fermion modes to be kept in the effective site, where $R \in SO(2L)$ and
\begin{equation}
Y_{P'} \equiv \mathop  \bigoplus \limits_{r = 1}^L \left[ {\begin{array}{*{20}c}
   0 & {g_r }  \\
   { - g_r } & 0  \\
\end{array}} \right],\quad g_r  = \left\{ {\begin{array}{*{20}c}
   1 & {r \le P'}  \\
   0 & {r > P'}  \\
\end{array}}. \right.\label{eq:P}
\end{equation}
Let $\Gamma_{3L}$ describe three consecutive blocks. Then the correlation matrix $\Gamma'_L$ for the effective site reads (Fig. (\ref{fig:disentanglers}))
\begin{equation}
\Gamma'_L = W^{\dagger}\left( {U \oplus U} \right)^{\dagger}\Gamma_{3L} \left( {U   \oplus U  } \right)W. \label{eq:truncated} 
\end{equation} 
Similarly, the correlation matrix $\bar{\Gamma}'_L$  for the modes to be removed is
\begin{eqnarray}
&&\bar{\Gamma}'_L  = \bar{W}^{\dagger}\left( {U \oplus U} \right)^{\dagger}\Gamma_{3L} \left( {U   \oplus U  } \right)\bar{W}, \label{eq:gone}\\ 
&&\bar{W} \equiv R(Y_L - Y_{P'}), ~~~~~~Y_L\equiv \oplus_{r=1}^L \left[ {\begin{array}{*{20}c}
   0 & 1  \\
   - 1 & 0  \\
\end{array}} \right]
\end{eqnarray} 

Our goal is to maximize the \emph{purity} of the modes to be projected out, so that they become as unentangled as possible. The sum of their purities, $\sum_{r=P'+1}^L v_r$, is half of the antisymmetric trace of $\tilde{\Gamma}'_L$, $\tr ( \tilde{\Gamma}'_L Y_L^{\dagger})$. Consequently, $U$ and $W$ are obtained from the optimization
\begin{equation}
 \max_{U,R\in SO(2N)} ~~\tr \left( \tilde{\Gamma}'_L Y_L^{\dagger}\right), \label{eq:optimization}
\end{equation}
that we address through a sequence of alternating optimizations for $U$ and $R$ \cite{optimizations}. 

\begin{figure}[!tb]
\begin{center}
\includegraphics[width=8cm]{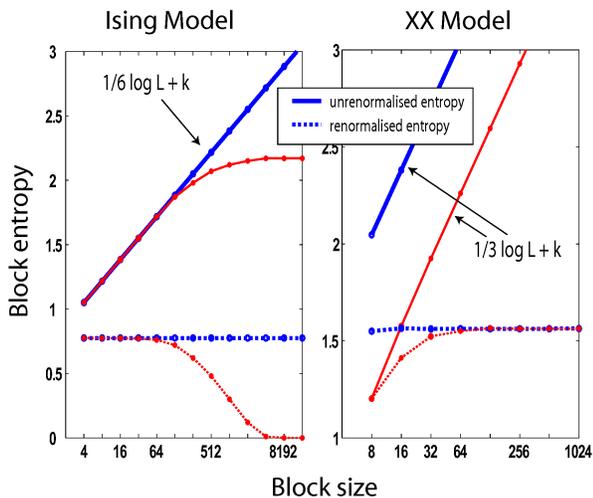}
\caption{ Scaling of the entanglement entropy $S_L$ \cite{logVidal} in 1D systems. \emph{Left:} Quantum Ising model, $\gamma=1$. Bold (solid/dotted) lines represent entanglement at criticality, $\lambda=1$. The system is an entangled fixed point of our RG transformation: the correlation matrices $\{\Gamma^{(1)}, \Gamma^{(2)}, \cdots \}$ quickly converge to a fixed $\Gamma_{\mbox{\tiny{Ising}}}^{*}$. In particular, the renormalized entanglement of a block is constant. Thin lines correspond to a non-critical system, $\lambda=1.001$, which the RG flow eventually brings a product (unentangled) ground state. \emph{Right:} Quantum XX model, $\gamma=0$. Bold/thin lines represent two critical cases, $\lambda=0$ and $\lambda=\textrm{cos} (15\pi/16)$. They belong to the same universality class and are found to indeed converge to the same correlation matrix $\Gamma_{\mbox{\tiny{XX}}}^{*}$, (with $\Gamma_{\mbox{\tiny{XX}}}^{*} \neq \Gamma_{\mbox{\tiny{Ising}}}^{*}$) and in particular to the same renormalized entropy. 
}\label{fig:1D}
\end{center} 
\end{figure}

Then, given the correlation matrix $\Gamma$ for $\ket{\Psi_{\mbox{\tiny GS}}}$, the RG transformation is implemented in three steps: ($i$) first a submatrix $\Gamma_{3L}$ for three consecutive blocks is extracted from $\Gamma$; ($ii$) then disentangler $U$ and isometry $W$ are computed using the optimization (\ref{eq:optimization}) while keeping $P'=P$ modes in the effective site; ($iii$) finally, $U$ and $W$ are used to transform the original $N$-mode system into a coarse-grained system with just $N/2$ modes and effective correlation matrix $\Gamma^{(1)}$. Some of the modes that are removed are still slightly mixed. Their mixness $\epsilon_r \equiv 1-v_r$ quantifies the errors introduced. Iteration of the RG transformation produces a sequence of increasingly coarse-grained lattices, described by correlation matrices $\{\Gamma^{(1)}, \Gamma^{(2)}, \cdots \}$. The corresponding disentanglers $\{U^{(1)}, U^{(2)}, \cdots\}$ and isometries $\{W^{(1)}, W^{(2)}, \cdots\}$ constitute the \emph{multi-scale entanglement renormalization ansatz} (MERA) \cite{MERA} for the ground state $\ket{\Psi_{\mbox{\tiny GS}}}$. 

\begin{figure}[tb]
\begin{center}
\includegraphics[width=8cm]{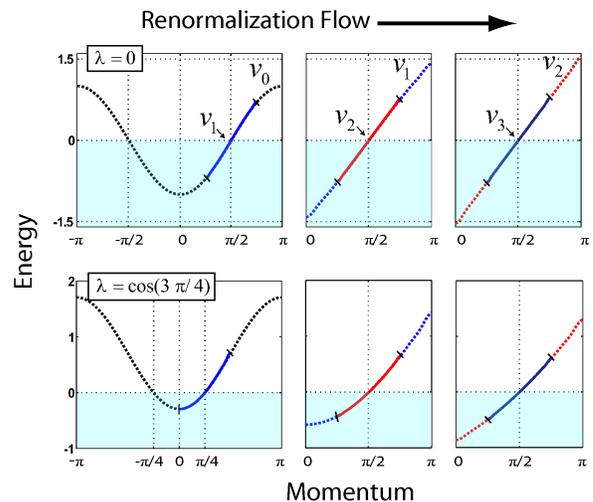}
\caption{Dispersion relation of Hamiltonian (\ref{eq:Ham}) in 1D with $\gamma=0$, quantum spin XX model, under successive RG transformations. Shading indicates the Fermi sea. A sequence of local, coarse-grained Hamiltonians is obtained $\{H^{(1)}, H^{(2)}, \cdots \}$ with their corresponding dispersion relations $\{\nu_1,\nu_2, \cdots\}$ converging to a straight line, a fixed point of the RG flow. Convergence is achieved very quickly at half filling ($\lambda=0$) and slower for $\lambda = \textrm{cos}(3\pi/4)$. These results have been obtained by minimizing the energy (Sect. IV of Ref.  \cite{algorithms}) while keeping $8$ modes in each effective site.} \label{fig:Ham}
\end{center} 
\end{figure}

\begin{figure}[!tb]
\begin{center}
\includegraphics[width=8cm]{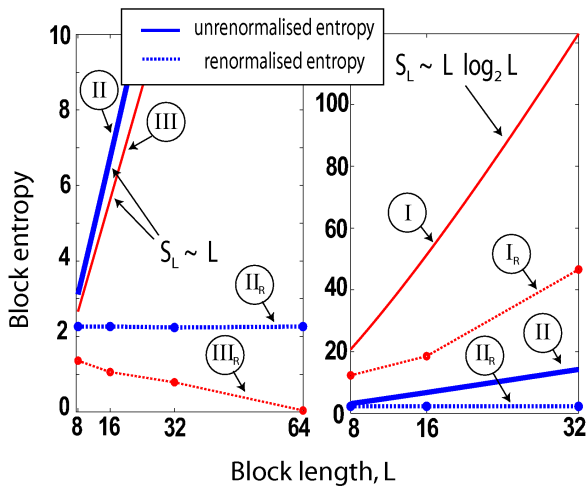}
\caption{Entanglement entropy $S_{L}$ of a block of $L\times L$ modes in 2D models. \emph{Left:} in the critical phase II and the non-critical phase III (bold/fine lines respectively) the entanglement entropy grows linearly with the size $L$ of the boundary of the block, $S_L \sim L$ (boundary law). As in 1D, the renormalized entanglement is constant for the critical model and it eventually vanishes for the non-critical model. We have considered $\gamma=1$ and $\lambda=2, \lambda=2.05$ for the critical/non-critical case. \emph{Right:} The critical phase II system $(\gamma,\lambda)=(1,2)$ is replotted for comparison against critical phase I, $(\gamma,\lambda)=(0,0)$, where the system has a 1D Fermi surface and the entanglement entropy has a logarithmic correction, $S_L\sim L\log L$. Here disentanglers are not able to reduce the renormalized entanglement down to a constant.} \label{fig:2D}
\end{center} 
\end{figure}

We have applied the present RG approach to Hamiltonian (\ref{eq:Ham}) in the thermodinamic limit $N\rightarrow \infty$. First we consider 1D systems, where the whole $(\gamma, \lambda)$ plane can be mapped into the quantum spin XY model using a Jordan-Wigner transformation. ($i$) For the line $\gamma=1$ [equivalent to the quantum spin Ising model] we consider $P=2$ modes per site and apply $13$ iterations of the RG transformation, so that a final effective site (with just $P=2$ modes) corresponds to $2\times 2^{13} = 16384$ modes of the original system. At the critical point $\lambda=1$, which is the most demaning, the mixness of the removed modes is at most $\epsilon_r = 1.2\times 10^{-4}$. The effect on local observables, even after the 13 iterations, is remarkably small: the error in the critical ground state energy is less than $10^{-7}$, while the two-point correlators $\left\langle {\hat a_r^\dag  \hat a_s} \right\rangle$, reconstructed from the MERA, accumulate a relative error that ranges from $10^{-7}$ for nearest neighbours to $10\%$ for $|r-s| \approx 4,000$. Had we not used disentanglers, the error in the energy would be $10^{-3}$ after only a single RG transformation and an error of $10\%$ in the two-point correlators is already achieved for $|r-s| = 42$. 
($ii$) The line $\gamma=0$ [equivalent to the quantum spin XX model] is critical for $|\lambda| < 1$. Here we consider $P=4$ modes per site and apply again 13 iterations of the RG transformation, reaching sizes of $4\times 2^{13} = 32768$ modes. The errors in energy and correlators are similar to those in the line $\gamma=1$. In both cases, an analysis of the RG flow and its fixed points in terms of entanglement is quite insightful, see Fig. (\ref{fig:1D}). ER can also be used to generate a RG transformation in the space of Hamiltonians, by replacing Eq. (\ref{eq:optimization}) with a minimization of the energy (see Sect. IV of Ref. \cite{algorithms} for details). Fig. (\ref{fig:Ham}) shows that critical systems are also fixed points of this alternative approach, that preserves the low energy spectrum.


In 2D the model has three phases, denoted I, II and III in Ref. \cite{Haas}, where the distinct forms of entanglement scaling were characterized. In phases II (critical, with a Fermi surface consisting of a finite number of points) and III (non-critical, with a gap in the energy spectrum) we are once more able to coarse-grain the system in a quasi-exact, sustainable manner. This is remarkable. The entropy of a square block made of $L^2$ modes grows as the size of its boundary, $S_L \sim L$ \cite{Haas}. This implies that the number of modes we should keep in an effective site grows \emph{exponentially} with the number of iterations of the RG transformation, which is precisely why DMRG does not work for large 2D systems. Instead, disentanglers bring this number again down to just a \emph{constant}. As a result one can, in principle, explore systems of arbitrary sizes. In particular, by considering $P = 4^2$ modes per site we apply $\tau =4$ iterations of the RG transformation, with a final block effectively spanning $P\times 4^{\tau+1}= 16384$ modes, whilst maintaining truncation errors of the same scale as the 1D models analysed, $\epsilon_r = 1.1\times 10^{-4}$. As in the 1D case, the structure of fixed points of the RG flow can be understood in terms of the renormalized entanglement, see Fig. (\ref{fig:2D}). On the other hand, Phase I (critical, with a one-dimensional Fermi surface) is so entangled that ER is no longer able to prevent the growth in the number $P'$ of modes that need to be kept per site. The system displays a logarithmic correction to the entropy, $S_L \sim L\log L$ \cite{Haas,Scholl,Wolf}, while the MERA can only reproduce a linear scaling $S_L\sim L$  \cite{MERA} if just a constant number of modes are kept per site, $P'=P$.

We have presented, in the simplified context of fermion models with quadratic Hamiltonian, unambiguous evidence of the validity of the ER approach in 1D and 2D systems. Similar derivations can be also conducted for bosonic lattice systems with quadratic Hamiltonians \cite{glen2}. Our results show, for the first time, that the MERA \cite{MERA} is an efficient description of 2D ground states. 
A number of examples also confirm that ($i$) ER produces a quasi-exact, real-space RG transformation where the coarse-grained lattice is locally equivalent to the original one, enabling the study of RG flow both in the space of ground states and Hamiltonians; ($ii$) non-critical systems end up in a stable fixed point of this RG flow, where the corresponding ground state is a product (i.e. fully disentangled) state, whereas critical systems end up in an unstable fixed point, with an entangled ground state. 

Interestingly, ER also sheds new light into the ground state structure of two-dimensional systems with a one-dimensional Fermi surface: only when logarithmic corrections appear for the entropy $S_L$ of a large block \cite{Haas,Scholl,Wolf}, does the above simple picture break down, suggesting the need for a generalized MERA \cite{glen2} in order to describe such systems.

The authors thank J. Fjaerestad and R. Orus for comments. Finantial support of the Australian Research Council (APA and FF0668731) is acknowledged. 


\textbf{Appendix A.---} Here we describe the process of coarse-graining a lattice by replacing blocks of sites with effective sites. We show that the truncation of the Hilbert space of a given block can be implemented by eliminating some of the modes in that block. 

We consider a fermionic lattice system in its (gaussian) ground state $\ket{\Psi_{\mbox{\tiny GS}}}$. Let $\mathbb{V}$ be the vector space of a block containing $L$ modes and let $\rho_{\mbox{\tiny GS}}$ denote the reduced density matrix of $\ket{\Psi}_{\mbox{\tiny GS}}$ on the block. We assume that the support of $\ket{\Psi}_{\mbox{\tiny GS}}$ is concentrated in a subpace $\mathbb{V}^{\rho} \subset \mathbb{V}$. Then, following White \cite{dmrg}, the optimal coarse-graining of the block is obtained by defining an effective site $s'$ with vector space $\mathbb{V}^{s'} = \mathbb{V}^{\rho}$. In our case, $\rho_{\mbox{\tiny GS}}$ is the tensor product of density matrices $\varrho_r$ for individual modes \cite{logVidal},
\begin{equation}
	\rho_{\mbox{\tiny GS}} = \bigotimes_{r=1}^L\varrho_{r} = \bigotimes_{r=1}^L \left( {\begin{array}{cc}
   \frac{1+v_r}{2} & 0  \\
   0 & \frac{1-v_r}{2}  \\
\end{array}} \right). \label{eq:rho2}
\end{equation}
Suppose that the first $P'$ modes are in a mixed state and the remaining $L-P'$ modes are in a pure state. Then we can write
\begin{equation}
	\rho_{\mbox{\tiny GS}} = (\bigotimes_{r=1}^{P'}\varrho_{r}) \otimes (\bigotimes_{r=P'+1}^{L}\varrho_{r}) \equiv \sigma \otimes \pi,
\end{equation}
where $\sigma$ is a mixed state with rank $2^{P'}$ whereas $\pi$ is a projector with rank 1. Let $\mathbb{V}= \mathbb{V}^{\sigma}\otimes \mathbb{V}^{\pi}$ be a tensor factorization of $\mathbb{V}$ such that $\sigma = \tr_{\mathbb{V}^{\pi}}(\rho)$. The key observation is that $\mathbb{V}^{\sigma} \cong \mathbb{V}^{\rho}$, and that $\rho$ and $\sigma$ have the same none-vanishing eigenvalues.
Therefore, we have two equivalent ways of constructing the space $\mathbb{V}^{s'}$ for the effective site $s'$ while preserving the support of the ground state density matrix $\rho_{\mbox{\tiny GS}}$. On the one hand, $\mathbb{V}^{s'}$ can be obtained by projecting $\mathbb{V}$ on the support $\mathbb{V}^{\rho}$ of $\rho$. On the other, $\mathbb{V}$ can also be build by factorizing the space $\mathbb{V}$ into two factor spaces $\mathbb{V}^{\sigma}$ and $\mathbb{V}^{\pi}$, and by then tracing out the second factor, corresponding to modes in a pure state. Both constructions lead to an equivalent effective lattice.
Finally, tracing out the factor space of mode $r$ corresponds, in the language of correlation matrices $\Gamma_L$, to removing the $r$th row and a $r$th column of $V\Gamma_L V^{\dagger}$ in Eq. (\ref{eq:GammaL}), process to which we referred to as \emph{projecting out} the mode.

\textbf{Appendix B.---} Here we describe the multi-scale entanglement renormalization ansatz (MERA) that we have used in the present work for 2D systems. It differs from the one described in Refs. \cite{ER,MERA}.

\begin{figure}[t]
\begin{center}
\includegraphics[width=8cm]{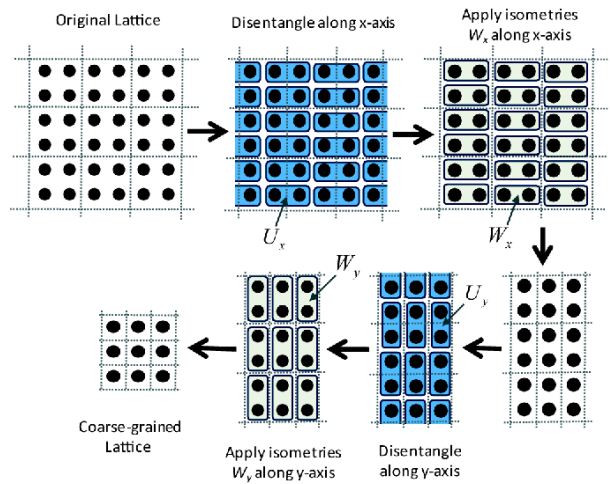}
\caption{2D MERA on a square lattice as described in \cite{ER,MERA}.}
\label{fig:MERA1}
\end{center} 
\end{figure}

The MERA can be understood as a peculiar class of quantum circuit with bounded size causal cones \cite{ER,MERA}. The causal cone structure of the circuit is the key property that allows for efficient computations with this ansatz. In practical realizations, the detailed local structure of the MERA (how disentanglers and isometries are interconnected) depends on the specific problem under consideration. In particular, for a 2D system the geometry of the lattice (square, triangular, Kagome,...) or the symmetries of the state that we intend to represent will be taken into consideration in order to choose a specific realization of the MERA. Fig. (\ref{fig:MERA1}) describes the 2D MERA for a square lattice discussed in \cite{ER,MERA}. Disentanglers and isometries act and reduce one lattice direction (say $x$ or $y$) at a time. This seems to be the most economical realization of a 2D MERA in terms of how the computation cost scales with the index dimension $\chi$.

\begin{figure}[!h]
\begin{center}
\includegraphics[width=6cm]{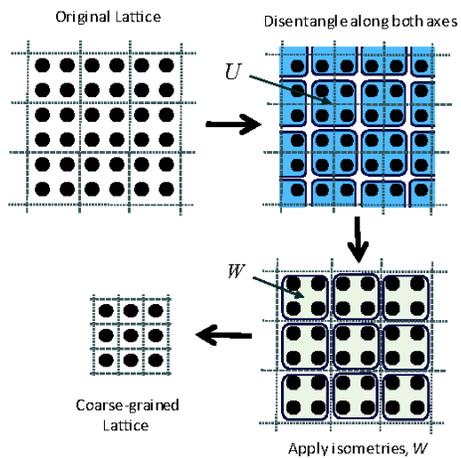}
\caption{2D MERA on a square lattice as used in the present calculations.}
\label{fig:MERA2}
\end{center} 
\end{figure}

Fig. (\ref{fig:MERA2}) describes instead the realization of the 2D MERA used in the present work. In this case both lattice directions are addressed simultaneously. While this realization of the MERA induces a cost that scales as a larger power of $\chi$, it seems to be more adequate for problems where it is important to preserve symmetry under $90^o$ rotations. Similarly, one could consider other realizations, provided the fundamental causal cone properties of the associated quantum circuit are preserved.

\emph{NOTE ADDED:} The 2D MERA realization used in this work has also been subsequently but independently discussed in Ref. \cite{Cincio}. Appendix B has been added to clarify how our present approach differs from the one in Refs. \cite{ER,MERA}.\

\begin{thebibliography}{99}
\bibitem{fisher} M.E. Fisher, Rev. Mod. Phys. {\bf 70}, 653 (1998).
\bibitem{kadanov} L. P. Kadanov, Physics {\bf 2}, 263 (1966).
\bibitem{dmrg} S. R. White, Phys. Rev. Lett. {\bf 69}, 2863 (1992), Phys. Rev. B {\bf 48}, 10345 (1993).
\bibitem{ER} G. Vidal, arXiv:cond-mat/0512165.
\bibitem{Haas} W. Li et al, arXiv:quant-ph/0602094
\bibitem{Scholl}  T. Barthel, M.-C. Chung, U. Schollwoeck,  Phys. Rev. A 74, 022329 (2006).
\bibitem{MERA} G. Vidal, quant-ph/0610099.
\bibitem{Wolf} M.M. Wolf, Phys. Rev. Lett. 96, 010404 (2006); D. Gioev, I. Klich, Phys. Rev. Lett. 96, 100503 (2006).
\bibitem{logVidal} G. Vidal et al, Phys. Rev. Lett. 90 (2003) 227902. E. Rico, J.I. Latorre, G. Vidal, Quant.Inf.Comput. 4 (2004) 48-92.  
\bibitem{algorithms} G. Vidal, arXiv:0707.1454
\bibitem{optimizations} Eq. (\ref{eq:optimization}) is quartic in $U$ and quadratic in $W$. We replace it with a series of linear optimizations along the lines of the strategies described in Sect. III of \cite{algorithms}.
\bibitem{glen2} G. Evenbly et al., in preparation. 
\bibitem{Cincio} L. Cincio, J. Dziarmaga and M. M. Rams, arXiv:0710.3829v1.

%
%
%
%
%
%
%
%
%
%
%
%
\end{thebibliography}
\end{document}